\documentclass[aip, reprint, ]{revtex4-2}

\usepackage{eurosym}

\usepackage[hidelinks]{hyperref}

\usepackage{csquotes}
\usepackage{etoolbox}
\usepackage{lmodern}

\usepackage{amsmath}
\usepackage{enumerate}
\usepackage{enumitem}
\usepackage{graphicx}
\usepackage{siunitx}[=v2]
\DeclareSIUnit{\pixel}{px}
\DeclareSIUnit{\fps}{fps}
\usepackage{float}
\usepackage[]{parskip} 
\usepackage[normalem]{ulem}

\usepackage{graphicx}
\usepackage{dcolumn}
\usepackage{bm}
\usepackage{tikz}
\usetikzlibrary{math}
\usetikzlibrary{shapes.geometric}
\usetikzlibrary{arrows}
\usetikzlibrary{positioning}
\usetikzlibrary{shapes}
\usetikzlibrary{intersections}
\usetikzlibrary{arrows.meta}
\usepackage{csvsimple}
\usepackage{changepage}
\usepackage[tbtags]{mathtools}
\usepackage{siunitx}
\usepackage{csquotes}

\usepackage{pgfplots}
\usepgfplotslibrary{groupplots}
\pgfplotsset{compat=newest}
\pgfplotsset{plot coordinates/math parser=false}

\usepackage[eulergreek]{sansmath}
\pgfplotsset{
  tick label style = {font=\sansmath\sffamily},
  every axis label = {font=\sansmath\sffamily},
  legend style = {font=\sansmath\sffamily},
  label style = {font=\sansmath\sffamily}
}

\makeatletter
\pgfplotsset{
    groupplot xlabel/.initial={},
    every groupplot x label/.style={
        at={($({group c1r\pgfplots@group@rows.west}|-{group c1r\pgfplots@group@rows.outer south})!0.5!({group c\pgfplots@group@columns r\pgfplots@group@rows.east}|-{group c\pgfplots@group@columns r\pgfplots@group@rows.outer south})$)},
        anchor=north,
    },
    groupplot ylabel/.initial={},
    every groupplot y label/.style={
            rotate=90,
        at={($({group c1r1.north}-|{group c1r1.outer
west})!0.5!({group c1r\pgfplots@group@rows.south}-|{group c1r\pgfplots@group@rows.outer west})$)},
        anchor=south
    },
    execute at end groupplot/.code={%
      \node [/pgfplots/every groupplot x label]
{\pgfkeysvalueof{/pgfplots/groupplot xlabel}};  
      \node [/pgfplots/every groupplot y label] 
{\pgfkeysvalueof{/pgfplots/groupplot ylabel}};  
    },
    group/only outer labels/.style =
{
group/every plot/.code = {%
    \ifnum\pgfplots@group@current@row=\pgfplots@group@rows\else%
        \pgfkeys{xticklabels = {}, xlabel = {}}\fi%
    \ifnum\pgfplots@group@current@column=1\else%
        \pgfkeys{yticklabels = {}, ylabel = {}}\fi%
}
}
}

\def\endpgfplots@environment@groupplot{%
    \endpgfplots@environment@opt%
    \pgfkeys{/pgfplots/execute at end groupplot}%
    \endgroup%
}
\makeatother

\DeclareSIUnit\Molar{\textsc{m}}
\DeclareSIUnit\rpm{rpm}

\newcommand*\chem[1]{\ensuremath{\mathrm{#1}}}

\usepackage[
	nonumberlist, 			
	acronym,      		
	nomain,		
	nopostdot,	
]{glossaries}
\usepackage{glossary-superragged}
\usepackage[hidelinks]{hyperref}
\usepackage{color, colortbl}

\usepackage{wrapfig}

\pgfplotsset{compat=newest}
\usepgfplotslibrary{fillbetween}
\usetikzlibrary{math}
\usetikzlibrary{calc}
\def\centerarc[#1](#2)(#3:#4:#5)
{ \draw[#1] ($(#2)+({#5*cos(#3)},{#5*sin(#3)})$) arc (#3:#4:#5); }

\usepackage{pgfplotstable}

\usetikzlibrary{external}
\usepackage{amssymb}

\usepackage{nicefrac}

\usepackage{multirow}
\usepackage{tabularx}
\usepackage{booktabs}
\usepackage{array}
\newcolumntype{L}[1]{>{\raggedright\let\newline\\\arraybackslash\hspace{0pt}}m{#1}}
\newcolumntype{C}[1]{>{\centering\let\newline\\\arraybackslash\hspace{0pt}}m{#1}}
\newcolumntype{R}[1]{>{\raggedleft\let\newline\\\arraybackslash\hspace{0pt}}m{#1}}

\newacronym{3d}{3D}{three dimensional}
\newacronym{am}{AM}{additive manufacturing}
\newacronym{fdm}{FDM}{fused deposition modeling}
\newacronym{ism}{ISM}{in-space manufacturing}
\newacronym{iss}{ISS}{International Space Station}
\newacronym{fcb}{FCB}{Functional Cargo Block}
\newacronym{dem}{DEM}{discrete element method}
\newacronym{md}{MD}{molecular dynamics}
\newacronym{dc}{DC}{direct-current}
\newacronym[plural=PFCs,firstplural=parabolic flight campaigns (PFCs)]{pfc}{PFC}{Parabolic Flight Campaign}
\newacronym{fft}{FFT}{Fast Fourrier Transform}
\newacronym{cad}{CAD}{Computer Assisted Design}
\newacronym{ptfe}{PTFE}{polytetrafluoroethylene}
\newacronym{ps}{PS}{polystyrene}
\newacronym{nasa}{NASA}{National Aeronautics and Space Administration}
\newacronym{esamm}{ESAMM}{Extended Structure Additive Manufacturing Machine}
\newacronym{amf}{AMF}{Additive Manufacturing Facility}
\newacronym{us}{US}{United States}
\newacronym{usa}{USA}{United States of America}
\newacronym{bmgs}{BMGs}{Bulk Metallic Glasses}
\newacronym{esa}{ESA}{European Space Agency}
\newacronym{si}{SI}{International System of Units, abbreviated from French \textit{Syst\`{e}me International (d'unit\'{e}s)}}
\newacronym{dlr}{DLR}{German Aerospace Center}
\newacronym{liggghts}{LIGGGHTS}{\acrshort{lammps} Improved for General Granular and Granular Heat Transfer Simulations}
\newacronym{lammps}{LAMMPS}{Large-scale Atomic/Molecular Massively Parallel Simulator}
\newacronym{sjkr}{SJKR}{Simplified Johnson-Kendall-Roberts}
\newacronym{ded}{DED}{Directed Energy Deposition}
\newacronym{slm}{SLM}{Selective Laser Melting}
\newacronym{sls}{SLS}{Selective Laser Sintering}
\newacronym{eva}{EVA}{Extra-Vehicular Activity}
\newacronym{sem}{SEM}{Scanning Electron Microscopy}
\newacronym{RPM}{RPM}{Ramdom Positioning Machine}
\newacronym{rpm}{rpm}{revolutions per minute}
\newacronym{rise}{RISE}{Research Internships in Science and Engineering}
\newacronym{daad}{DAAD}{German Academic Exchange Service, abbreviated from German \textit{Deutscher Akademischer Austauschdienst}}
\newacronym{fsm}{FSM}{finite-state machine}
\newacronym{ir}{IR}{infrared}
\newacronym{pcbs}{PCBs}{Printed Circuit Boards}
\newacronym{pcb}{PCB}{Printed Circuit Board}
\newacronym{mcr}{MCR}{Modular Compact Rheometer}
\newacronym{sff}{SFF}{Solid Freeform Fabrication}
\newacronym{uv}{UV}{ultraviolet}
\newacronym{abs}{ABS}{acrylonitrile butadiene styrene}
\newacronym{hpde}{HPDE}{high density polyethylene}
\newacronym{pei}{PEI}{polyetherimide}
\newacronym{bff}{BFF}{BioFabrication Facility}
\newacronym{lens}{LENS}{Laser Engineered Net Shaping}
\newacronym{cnc}{CNC}{Computer Numerical Control}
\newacronym{ebf3}{EBF$^3$}{Electron Beam Free-Form Fabrication}
\newacronym{leo}{LEO}{Low Earth Orbit}
\newacronym{pc}{PC}{polycarbonate}
\newacronym{crissp}{CRISSP}{Customisable Recyclable International Space Station Packaging}
\newacronym{Athena}{Athena}{Advanced Telescope for High-ENergy Astrophysics}
\newacronym{lbm}{LBM}{Laser Beam Melting}
\newacronym{bam}{BAM}{Federal Institute for Materials Research and Testing, abbreviated from German \textit{Bundesanstalt f\"{u}r Materialforschung und-pr\"{u}fung}}
\newacronym{pbf}{PBF}{powder bed fusion}
\newacronym{eb}{EB}{Electron Beam}
\newacronym{2d}{2D}{two dimensional}
\newacronym{4d}{4D}{four dimensional}
\newacronym{ft4}{FT4}{Freeman Technology 4 Powder Rheometer}
\newacronym{dsc}{DSC}{Differential Scanning Calorimetry}
\newacronym{pmma}{PMMA}{polymethylmethacrylate}
\newacronym{1g}{$1g$}{gravity on-ground}
\newacronym{mug}{$\mu g$}{microgravity}
\newacronym{bcm}{BCM}{Box Counting Method}
\newacronym{mct}{MCT}{Mode Coupling Theory}
\newacronym{gmct}{gMCT}{granular Mode Coupling Theory}
\newacronym{itt}{ITT}{Integration Through Transients}
\newacronym{mfc}{MFC}{Mass Flow Controller}
\newacronym{ct}{CT}{computed tomography}
\newacronym{xct}{XCT}{X-ray computed tomography}
\newacronym{cv}{CV}{curriculum vitae} 
\newacronym{pi}{PI}{principal investigator}
\newacronym{osp}{OSP}{orthogonal superimposed perturbation}
\newacronym{npi}{NPI}{Network Partnering Initiative}
\newacronym{ecsat}{ECSAT}{European Centre for Space Applications and Telecommunications}
\newacronym{eac}{EAC}{European Astronaut Centre}
\newacronym{estec}{ESTEC}{European Space Research and Technology Centre}
\newacronym{fps}{fps}{frames per second}
\newacronym{pdf}{pdf}{probability density function}
\newacronym{al}{Al}{aluminium}
\newacronym{ss}{\textit{SS}}{\textit{Smooth Surface}}
\newacronym{rs}{\textit{RS}}{\textit{Rough Surface}}
\newacronym{rcp}{rcp}{random close packing}
\newacronym{iop}{IoP UvA}{Institute of Physics of the University of Amsterdam}
\newacronym{mp}{MP}{Institute of Material Physics for Space}
\newacronym{elgra}{ELGRA}{European Low Gravity Research Association}
\newacronym{zarm}{ZARM}{Center of Applied Space Technology and Microgravity}
\newacronym{piv}{PIV}{particle image velocimetry}
\newacronym{ptv}{PTV}{particle tracking velocimetry}
\newacronym{oct}{OCT}{optical coherence tomography}
\newacronym{ctab}{CTAB}{cetyltrimethylammoniumbromide}
\newacronym{nasal}{NaSal}{sodium salicylate}
\newacronym{cpcl}{CPCl}{cetylpyridinium chloride}
\newacronym{ccs}{CCS}{capsule control cystem}
\newacronym{cw}{CW}{continuous-wave}
\newacronym{dpss}{DPSS}{diode-pumped solid-state}
\newacronym{tc}{TC}{Technical Center}
\newacronym{paa}{PAA}{poly acrylic acid}
\newacronym{ixa}{IXA}{Innovation Exchange Amsterdam}
\newacronym{nwo}{NWO}{Dutch Research Council} 
\newacronym{pp}{PP}{polypropylene}
\newacronym{pu}{PU}{polyurethane}

\usepackage{filecontents}

\begin{document}

\title{\textbf{
Spreading of droplets under various gravitational accelerations
}}

\author{Olfa D'Angelo}
\email[]{Author to whom correspondence should be addressed: olfa.dangelo@mail.com}
\affiliation{Institute for Multiscale Simulation, Universit\"{a}t Erlangen-N\"{u}rnberg, Cauerstra\ss{}e 3, 91058 Erlangen, Germany}
\affiliation{Institute of Materials Physics in Space, German Aerospace Center (DLR), Linder H\"{o}he, 51170 Cologne, Germany}

\author{Felix Kuthe}
\affiliation{Cologne Lab for Artificial Intelligence and Smart Automation, University of Applied Science Cologne, Betzdorfer Stra{\ss}e 2, 50679 K\"{o}ln, Germany}
\affiliation{Institute of Materials Physics in Space, German Aerospace Center (DLR), Linder H\"{o}he, 51170 Cologne, Germany}

\author{Kasper van Nieuwland}
\affiliation{Technology Centre, University of Amsterdam, Science Park 904, 1098 XH Amsterdam, The Netherlands}

\author{Clint Ederveen Janssen}
\affiliation{Technology Centre, University of Amsterdam, Science Park 904, 1098 XH Amsterdam, The Netherlands}

\author{Thomas Voigtmann}
\email[]{thomas.voigtmann@dlr.de}
\affiliation{Institute of Materials Physics in Space, German Aerospace Center (DLR), Linder H\"{o}he, 51170 Cologne, Germany}
\affiliation{Institute for Theoretical Physics, Heinrich-Heine-Universit\"at D\"usseldorf, Universit\"atsstra{\ss}e 1, 40225 D\"usseldorf, Germany}

\author{Maziyar Jalaal}
\email[]{m.jalaal@uva.nl}
\affiliation{Van der Waals-Zeeman Institute, University of Amsterdam, Science Park 904, 1098 XH Amsterdam, The Netherlands}

\date{\today}

\begin{abstract}
We describe a setup to perform systematic studies on the spreading of droplets of complex fluids under microgravity conditions.
Tweaking the gravitational acceleration under which droplets are deposited provides access to different regimes of the spreading dynamics, quantified through the Bond number. In particular, microgravity allows to form large droplets while remaining in  the regime where surface tension effects and internal driving stresses are predominant over hydrostatic forces.
The \textsc{vip-drop\textsuperscript{2}} (\textsc{vi}sco-\textsc{p}lastic \textsc{drop}lets on the \textsc{drop} tower) experimental module
provides a versatile platform to study a wide range of complex fluids through
the deposition of axisymmetric droplets.
The module offers the possibility to deposit droplets on a precursor layer,
which can be composed of the same or of a different fluid.
Besides, it allows to deposit four droplets simultaneously, while conducting shadowgraphy on all of them, and observing either the flow field (through \acrlong{piv}), or the stress distribution inside the droplet in the case of stress birefringent fluids.
Developed for a drop tower catapult system, it
is designed to withstand a vertical acceleration of up to 30 times Earth's gravitational acceleration in the downwards direction, and can operate remotely, under microgravity conditions.
We provide a detailed description of the module, 
and exemplary data analysis for droplets spreading on-ground and in microgravity.
\end{abstract}

\pacs{}

\maketitle

\definecolor{brickred}{rgb}{0.8, 0.25, 0.33}
\definecolor{darkorange}{rgb}{1.0, 0.55, 0.0}
\definecolor{persiangreen}{rgb}{0.0, 0.65, 0.58}
\definecolor{persianindigo}{rgb}{0.2, 0.07, 0.48}
\definecolor{cadet}{rgb}{0.33, 0.41, 0.47}
\definecolor{turquoisegreen}{rgb}{0.63, 0.84, 0.71}
\definecolor{sandybrown}{rgb}{0.96, 0.64, 0.38}
\definecolor{blueblue}{rgb}{0.0, 0.2, 0.6}
\definecolor{ballblue}{rgb}{0.13, 0.67, 0.8}
\definecolor{greengreen}{rgb}{0.0, 0.5, 0.0}
\definecolor{bittersweet}{rgb}{1.0, 0.0, 0.5}

\definecolor{myX}{RGB}{102,194,165}
\definecolor{myY}{RGB}{252,141,98}
\definecolor{myZ}{RGB}{141,160,203}

\definecolor{c1}{rgb}{0.7068574918274737, 0.11027871818526241, 0.2747061222663145}%
\definecolor{c2}{rgb}{0.6780437401750237, 0.1857033496010309, 0.10475664313058389}%
\definecolor{c3}{rgb}{0.5819819292481591, 0.28135510723834917, 0.10365807489974799}%
\definecolor{c4}{rgb}{0.5234892407222805, 0.3183040932830017, 0.10308831855626377}%
\definecolor{c5}{rgb}{0.4801916866157751, 0.3399605081271155, 0.10271364194308724}%
\definecolor{c6}{rgb}{0.44359832760663015, 0.3553988979685888, 0.10242748858902176}%
\definecolor{c7}{rgb}{0.40913672397205286, 0.36794412723413256, 0.10218299803254591}%
\definecolor{c8}{rgb}{0.37320211354661986, 0.37925108836394167, 0.10195327596798023}%
\definecolor{c9}{rgb}{0.33135856975649947, 0.3904277497779026, 0.10171733001337946}%
\definecolor{c10}{rgb}{0.2751398809612443, 0.40252912490464393, 0.10145175401497963}%
\definecolor{c11}{rgb}{0.17705545280426335, 0.4169982861326329, 0.10112016988851782}%
\definecolor{c12}{rgb}{0.10370113411519366, 0.41987331701008007, 0.20784080256789766}%
\definecolor{c13}{rgb}{0.10672620229256541, 0.415683426104876, 0.2824387977867041}%
\definecolor{c14}{rgb}{0.10896052083074859, 0.412482459879469, 0.3262929409150867}%
\definecolor{c15}{rgb}{0.11082049183583692, 0.4097465781098524, 0.3585658834499722}%
\definecolor{c16}{rgb}{0.1125289398766243, 0.40717495121262287, 0.38576967807315987}%
\definecolor{c17}{rgb}{0.11424591411818866, 0.4045324469356048, 0.41127417284765605}%
\definecolor{c18}{rgb}{0.11613475317261168, 0.4015563987219263, 0.4376207984793854}%
\definecolor{c19}{rgb}{0.11843118069754927, 0.3978377246048391, 0.46769849466738594}%
\definecolor{c20}{rgb}{0.1215889511379443, 0.3925371130865723, 0.5062751486798138}%
\definecolor{c21}{rgb}{0.12675667510032929, 0.38336665224260497, 0.564172135389151}%
\definecolor{c22}{rgb}{0.13823697094516182, 0.36050804998003744, 0.6775165660716256}%
\definecolor{c23}{rgb}{0.2929229731919379, 0.2517707788576924, 0.9106622939003164}%
\definecolor{c24}{rgb}{0.5044517963791875, 0.15823641315504755, 0.8374462940274711}%
\definecolor{c25}{rgb}{0.5743505600217842, 0.14563681653078617, 0.7277475942695497}%
\definecolor{c26}{rgb}{0.6114384687170589, 0.13753415571759905, 0.6502354669913063}%
\definecolor{c27}{rgb}{0.6363457404298491, 0.13141740987926892, 0.586320806947322}%
\definecolor{c28}{rgb}{0.6557132733878439, 0.12622710616005786, 0.5268534496956088}%
\definecolor{c29}{rgb}{0.6725336221941806, 0.12137095721249477, 0.4649001891758376}%
\definecolor{c30}{rgb}{0.688593421429631, 0.11639483989072555, 0.39157406584043325}%

\section{Introduction}

The deposition and spreading of complex fluids on a solid surface 
is at the basis of numerous industrial applications:
\gls{am}~\cite{placone2018recent, kyle2017printability, jiang2020extrusion, buswell20183d},
various coating methods~\cite{weinstein2004coating, ruschak1985coating}, 
inkjet printing~\cite{derby2010inkjet, lohse2022fundamental},
and other technical processes involving 
heat transfer~\cite{Yonemoto2017}
or energy harvesting~\cite{Ueda1979}.
The spreading of thin films, filaments or droplets (from Newtonian to polymeric fluids) has extensively been studied to improve and optimize the current technologies~\cite{lu2016critical, rafai2005spreading, bonn2009wetting, bergeron2000controlling, Jalaal2020a}. Unsurprisingly, most previous studies have been performed under Earth's gravitational acceleration; 
knowledge on spreading under low- and microgravity is limited,
although it would be beneficial not only for the development of space technologies, 
but also
as a tool for understanding phenomena that happen on-ground.
This knowledge gap is even more pronounced for fluids with complex rheological properties, such as viscoelasticity and viscoplasticity.
We present an experimental setup 
to address
this knowledge gap.

The spreading of fluids is typically a function of surface tension, the interaction with the solid surface, inertia, body forces (including gravity), and the rheological properties of the material.
In this context, the importance of gravity is typically measured by comparing the hydrostatic pressure $\rho g \mathcal{L}$ (where $\rho$ is the fluid's density, $g$
the gravitational acceleration, and $\mathcal L$ a typical length scale
of the droplet) and the capillary pressure $\sigma/\mathcal{L}$ (where $\sigma$ is the surface tension of the fluid). The relevant dimensionless
number is the Bond number,
\begin{equation}\label{eq:bo}
                \mathcal{B} = \frac{\rho \, g \, \mathcal{L}^2}{\sigma}\,,
\end{equation}
where both $\rho$ and $\sigma$ typically cannot be changed significantly
without also modifying other material properties.  
A plethora of experimental work addresses the regime of large $\mathcal B$,
which represents the gravity-dominated regime, applicable to large-scale
phenomena on ground, such as landslides or lava flows \cite{Griffiths.2000,Bassis.2019}. Our focus is however the limit of $\mathcal{B} \rightarrow 0$: 
the regime of interest for technological applications relying on the deposition of \emph{very small} droplets ($\mathcal{L} \rightarrow 0$, typically relevant for sprays, coatings, or printing), but also for the spreading of finite size structures under low and microgravity
($g \rightarrow 0$).

Investigating the spreading of complex fluids under various 
gravitational accelerations
is important for at least two reasons. First, such studies are crucial for developing space technologies that rely on the spreading of complex fluids. Examples include
\gls{am}
of thermo-plastics (e.g., for on-site printing tools in space stations), 2D and 3D printing soft materials (e.g., high precision electronics and bioprinting of food, organs or living tissues), and large-scale 3D printing of cement-like materials for possible habitat 
on the Moon or Mars~\cite{wong20143d, joshi20153d, dunn20103d, leach20143d, Makaya2022}.
Secondly, experiments under low gravity allow exploring limits that are difficult to study under the Earth's gravity.
For instance, on ground, the limit of negligible gravitational effects is mainly achievable by significantly reducing the characteristic size of the droplets (i.e.~$\mathcal{L} \to 0$). 
This, however, poses experimental intricacy because of boundary effects,
e.g.~from the details of the deposition procedure and perturbations imposed by the deposition nozzle.
Imaging techniques can also impose resolution limitations,
and thus the determination of internal velocity and stress fields becomes challenging in small droplets. Hence, experiments on the spreading of complex fluids under low gravity also shed light on our general understanding of the physics of spreading of soft matter and complex fluids on Earth.

Gravity-related experimental platforms such as drop towers~\cite{Dittus1991, VonKampen2005}, parabolic flights~\cite{Pletser2016} or sounding rockets~\cite{Pallone2018}, among others, are great candidates to provide insight into the 
spreading of complex fluids
under different gravitational accelerations.
Here
we describe a setup designed to
allow for the investigation of the spreading of droplets (with controlled volume and
deposition speed) in weightlessness ($g \rightarrow 0$).
Studies of physical characteristics of droplets 
can be highly sensitive to gravitational acceleration
perturbations ($g$-jitter), such as that due to parabolic flight
environments;
we take advantage of the high quality of microgravity 
offered by drop towers to  
minimize such perturbations.
The experimental hardware was developed for the \gls{zarm}
drop tower \cite{Dittus1990,Dittus1991,VonKampen2005} (Bremen, Germany)
under the name \textsc{vip-drop\textsuperscript{2}} (\textsc{vi}sco-\textsc{p}lastic \textsc{drop}lets on the \textsc{drop} tower).

Axisymmetric droplets
provide an excellent demonstration system
towards understanding and modeling the spread of complex fluids. 
On the one hand, they are closely related to applications where single droplets or filaments are deposited on a surface (e.g., printing). On the other hand, despite its geometrical simplicity, axisymmetric spreading remains a
complex problem because it involves dynamic phenomena happening simultaneously on multiple length scales.
The spreading of Newtonian droplets has been studied 
in various regimes~\cite{Bergemann2018},
including under microgravity~\cite{Diana2012,Brutin2009,Abel2004,Ababneh2006, Kabov2013}:
experimental work on the effect of gravity on droplets' morphology and spreading 
mainly focused on Newtonian fluids spreading on a solid surface.
Studies on droplets of non-Newtonian fluids under weightlessness  
have hitherto prioritized the evaporation dynamics of 
complex fluids droplets~\cite{Carle2013, Kumar2020, Li2020}
and the solidification of metal droplets \cite{Huang2021}
towards droplet-based printing.

Our experimental setup can be used for materials with various rheological properties. 
A main focus of our work is to study viscoplastic fluids
(also known as yield stress fluids)~\cite{bonn2017yield, balmforth2014yieldin}. Such materials behave like an elastic solid at low stresses, but above a critical stress (the yield stress),
flow like a viscous fluid --
a property that makes them highly desirable in many technical and technological applications.
The spreading of viscoplastic fluids features a finite spreading time, 
as it ceases
when stress everywhere inside the droplet is below the yield stress~\cite{Jalaal2020a}. 
Another class of fluids that we study are shear-thinning viscoelastic fluids~\cite{Amador2016, Chen2022, Gladden2010},
characterized by a large structural relaxation time.
Such material behaves as a fluid 
with an apparent yield stress on time scales
that are short compared to their relaxation time.

The setup is inherently modular,
in that it allows for a wide variety of fluids to be tested,
with no predetermined limitation in yield stress or viscosity of the experimental fluid.
The module allows to study up to four droplets in parallel, in order to
optimize microgravity time.
As shown by \citeauthor{Brutin2009}~\cite{Brutin2009} and \citeauthor{Diana2012}~\cite{Diana2012}, 
the gravitational acceleration during droplet deposition 
influences the droplet's shape,
during and after spreading. 
In contrast to previous work,
deposition and spreading are conducted 
under the same gravitational acceleration,
ensuring 
that spreading dynamics occurs under constant 
conditions.
Besides, the module offers the possibility to simultaneously
conduct three types of measurement on all or some of the droplets:
simple shadowgraphy provides basic information on the
temporal evolution of the spreading;
\gls{piv} on fluids with embedded tracer particles
provides simultaneous information on the internal flow fields;
internal stresses can be visualized by a rheo-optical method for
fluids that show flow-induced birefringence.

Another important characteristic of the experimental module is that
the deposition can be done directly on a substrate of a chosen material,
but can also be preceded by the deposition of a thin film of variable height,
constituted of either the same or a different fluid.
As surface-specific interactions complicate matters for the spreading of
droplets, the problem can be simplified by studying
droplets on a pre-wetted surface.
The application of such a precursor layer can also produce
a more appropriate representation of the system of interest
(for example, in \gls{am} and paint
spray applications, liquid droplets or filaments are deposited on an existing layer of the fluid).

In the following, we describe the design and implementation of the
\textsc{vip-drop\textsuperscript{2}} module, and first reference measurements
that have been obtained during three drop-tower campaigns at \gls{zarm} in the course of the years 2021 and 2022.
We first detail the available measurement techniques that are
implemented in the module (Sec.~\ref{sec:techniques}),
then describe the specifications and implementation-specific
details of the module (Sec.~\ref{sec:specifications}).
Initial experimental results are presented in Sec.~\ref{sec:results},
after which a summary and outlook are presented in Sec.~\ref{sec:conclusion}.

\section{Measurement Methods}\label{sec:techniques}
Three measurement techniques are implemented
in the \textsc{vip-drop\textsuperscript{2}} module.
We start with a brief summary of the principles of each of those techniques.

\subsection{Shadowgraphy}
A simple but essential parameter
to describe
the spreading of a droplet is its height profile $h(t)$ as a
function of time. Shadowgraphy is a robust and easy experimental technique
to gather this information:
illuminating the droplet from one side with a homogeneous light source,
and observing it with a camera from the opposite side, images are obtained
where the radial extent of the droplet is visible. In principle, since
inhomogeneities in optical media change the light refraction, shadowgraphy
gives access to more detailed information, but in the present situation
the binary information encoding the droplet cross-section is sufficient.
A simple edge-detection algorithm can thus be employed to process the images
in order to extract the time-dependent droplet profile, under assumption of
axial symmetry around the depositing nozzle.

From the profile $h(t)$, we can in particular extract the droplet radius
$R(t)$. In the case of yield-stress fluids whose spreading comes to a
halt at a finite radius, the long-time behavior of $R(t)$ allows to determine
the asymptotic radius $R_f$ for which theoretical predictions in the
form of separate scaling laws for the
regimes $\mathcal B=0$ and $\mathcal B\gg1$ exist \cite{Jalaal2020a}.
Another quantity of interest that can be extracted from the shadowgraphy
images is the (apparent) contact angle of the droplet with the substrate.

Shadowgraphy is also possible from below the droplet, which gives access
to the time-dependent shape of the droplet's rim, and thus allows to explicitly
check the assumption of axisymmetry.

\subsection{\Acrlong{piv}}\label{measurement_piv}

Of specific interest beyond the information on the macroscopic shape
of the droplet, is its internal flow field.
Calculations based on the Navier-Stokes equations in the
thin-film limit and supplemented with empirical material laws,
suggest that the flow field of a yield-stress fluid has a rich phenomenology:
owing to the fact that locally the imposed stress remains below the yield
stress of the fluid, a stagnation zone appears in the center, and a plug-flow
zone near the top surface of the droplet, separated by a localized flow
boundary \cite{Jalaal2020a}.

We employ \acrfull{piv} \cite{Adrian1991,Westerweel2013,RaffelPIV}
to observe the flow field of the spreading droplets.
\Gls{piv} gives access to the Eulerian velocity field, i.e., the locally
resolved instantaneous fluid velocity.
For this, flourescently labelled tracer particles (typically with sizes
in the range of \SIrange{1}{10}{\micro\meter}) are embedded in the fluid,
small enough to provide least possible disturbance to the flow field, and
large enough to be visible, and such that they can be assumed to be advected
by the fluid flow. A laser sheet illuminates one cross-section of
the sample, so that the flourescent emission of the tracers in this
cross-section can be detected by a camera. From the analysis of two consecutive
frames
captured by a high-speed camera, the local velocity of the flow at the position
of the tracers can be determined.

\Gls{piv} has been employed previously for drop-spreading experiments
\cite{Jalaal2015}, albeit observing through an inverted microscope and a
confocal scanning unit placed below the droplet. In the present setup,
the laser sheet is oriented such that a cross-sectional plane orthogonal to
the surface of deposition is observed with a camera placed on the side of
the droplet. This gives information on the height-dependence of the velocity
field; assuming that the flow inside the droplet is axisymmetric to a good
approximation, the full information can then be reconstructed, if the laser
sheet is placed to cut through the central axis of the droplet as closely
as possible.

\subsection{Rheo-optical Measurements}

Flow in fluids in general causes internal stresses. In particular
in non-Newtonian fluids, the transient evolution of these stresses as
a function of time does not instantaneously follow the current flow field,
but stresses build up and are released with a time delay that depends on
the local structural relaxation mechanisms in the material.
Spatially resolved measurements of internal stresses in the droplets
thus reveal
localized rearrangements in the structure of the fluid as it spreads.

A convenient method to observe internal stresses in optically transparent
materials, at least qualitatively -- in specific cases, also quantitatively --
is based on the stress-optical law first formulated by Maxwell:
it links the optical indices of refraction to the principal stresses
inside the material.
For solid materials, the observation of the resulting load-dependent
birefringence patterns is a technique called \emph{photoelasticity}.
Since the classical work of
Frocht~\cite{Frocht}, it has
developed into a standard method for the assessment of internal stresses
in transparent solids or transparent analog-models of mechanical parts~\cite{Ramesh2000}.

In fluids, the transient stresses, by the same physical mechanism, give
rise to a phenomenon called flow-induced birefringence. It is
a powerful optical rheological method used to obtain information
on the flow of complex fluids, especially in polymer melts which
display a large rheo-optical effect~\cite{Gladden2010, Soulages.2008}.

The principle of rheo-optics is the assumption that at each point
within the material, the optical properties are expressed in terms of
three principal refractive indices $n_i$ ($i=1,2,3$)
for electromagnetic waves whose
polarization is aligned with the principal stresses $\tau_i$,
\begin{equation}\label{eq:stress-optical-law}
  n_i-n_j=C(\tau_i-\tau_j)\,,
\end{equation}
with a material-specific constant $C$ called the stress-optic or
photoelastic constant.
Hence, the material becomes optically birefringent in response to stress,
such that transmitted polarized light is rotated by an angle that corresponds
to the distribution of local stresses along the light-ray's path through
the material. Observing the transmitted light under crossed polarizers
thus results in typical birefringence patterns that can be analyzed.
In practice, the use of circular (instead of linear) polarizers helps to
eliminate further spurious transmission from refraction~\cite{Zadeh2019}
and eliminates dark lines known as isoclinics that arise purely from
the geometry of the linear-polarizer setup.
Taking images both from the side and from below the droplet,
two different tomographic projections of the internal stresses can be
obtained.

\section{Specifications}\label{sec:specifications}

\subsection{Microgravity platform}

The \textsc{vip-drop\textsuperscript{2}} module is designed for the \gls{zarm} drop tower (see Fig.~\ref{fig:exp_full}) situated in Bremen, Germany~\cite{Dittus1990, Dittus1991, VonKampen2005}. 
This drop tower offers the possibility to perform experiments in weightlessness
by letting the experiment fall for approx.~\SI{4.6}{\s} in an
evacuated tube,
hence avoiding aerodynamic drag to achieve true free-falling conditions.
The available time for such \enquote{flight} can be
essentially doubled by first catapulting the experiment module to the top of the
tower. Under these conditions, the experiment module is effectively
in microgravity for approx.~\SI{9.3}{\s}, with a residual acceleration smaller than $\approx 10^{-4} g$
(where $g \approx \SI{9.81}{\meter\per\second\square}$ is Earth's gravitational acceleration).

The high quality of microgravity that is obtained in the \gls{zarm} drop tower allows to conduct experiments that are very sensitive to gravity-jitters.
This is important in droplet-spreading experiments in order to avoid
possible effects of vibrations and gravity jitters on the spreading~\cite{Garg2021}.

\subsection{VIP-DROP\textsuperscript{2} module}

\begin{figure*}
	\centering
    \includegraphics{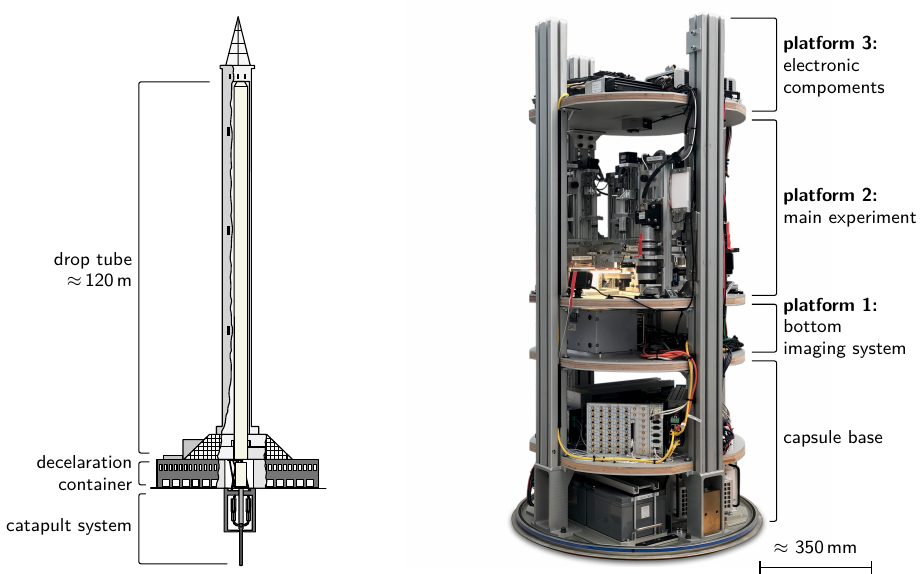}
	\caption{\label{fig:exp_full}
Schematic of the \gls{zarm} drop tower (left), and full VIP-DROP\textsuperscript{2} module (right), composed of three experiment platforms and the capsule base.
The capsule base houses the control system for drop-tower operations and the battery.
  Platform~1 contains the high-speed camera recording systems and the camera
  heads for observations of the droplets from below.
  Platform~2 contains the main experiment: the droplet deposition areas
  mounted on a rotary stage, syringe systems for pre-wetting and deposition,
  and backlighting LED panels. Side-view cameras and a laser unit for
  \gls{piv} are mounted vertically on the stringers.
  Electronic components on platform~3 include control computers and
  the laser driver.
}
\end{figure*}

The experimental module, shown in Figure~\ref{fig:exp_full}, 
is divided into three platforms arranged vertically and mounted on
stringer elements forming the base structure of the drop-tower capsule. The
intermediate platform (platform~2) houses the main experiment.
Underneath, platform~1 contains the recording modules and bottom imaging systems (to observe the droplets from below);
electronic components are mounted on platform~3, situated on top of the module.
The module is made to attach to the capsule-base from \gls{zarm},
which contains the \gls{ccs} and additional elements, such as sensor systems and battery. 
Covered by an outer shell, 
this assembly forms the final capsule
launched in the drop tower.

\begin{figure*}
	\centering
	\includegraphics{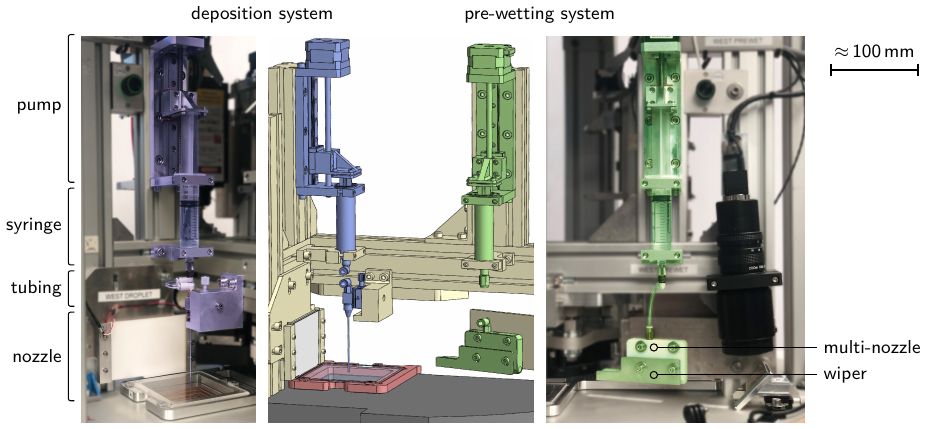}
	\caption{Structure of one section of the main experiment platform: \gls{cad} model (center panel), and close-ups of the deposition and pre-wetting system (respectively left and right); the glass substrate in experiment position is also visible.}
	\label{fig:section_overview}
\end{figure*}

The simultaneous deposition of four droplets is made possible by
dividing the main experiment platform into four equal sections.
Each section allows for the independent preparation and examination of
one droplet.
Each droplet is deposited on a glass substrate
(quartz glass plates from proQuarz GmbH)
that can be pre-wetted just before the droplet deposition.
Each section is equipped with separate custom-made pump systems for both the
deposition and the pre-wetting,
so that four different combinations of droplet and (optionally different)
pre-wetting material can be examined simultaneously during one flight.
A single deposition system (to deposit the droplet) and the associated pre-wetting system 
are shown in Figure~\ref{fig:section_overview}.

All pumping systems are controlled individually,
which allows to deposit different quantities of liquid 
at different rate per position. 
The motor control also implements a configurable amount of retraction
of the syringe pistons, in order to stop the deposition of
yield-stress fluids as instantaneously as possible: without retraction,
the cessation of the flow out of the deposition nozzle produces some leakage
that depends on the rheological properties of the fluid.
The syringes currently used are \SI{20}{\milli\liter} syringes made of \gls{pp} from Braun Omnifix;
connections are made with \gls{pu} tubing of diameter \SI{4}{\mm} from SMC Corporation.
A purging feature can be implemented (for example, right before launch)
to provide clean conditions at the start of the deposition:
in the prolonged waiting time prior to experiment start, small quantities
of fluid can dry at the tip of the nozzle,
or air bubbles can appear in the system.
A small recipient is placed on the side on the substrate holder (i.e.~attached to the rotary stage; not shown in Fig.~\ref{fig:section_overview}) in order to retrieve the purged material.

The theoretical resolution of the fluid dispensing is one micro-step of the motor driving the pump,
which represents a volume on the order of $10^{-7}$\,\si{\liter}.
In practice, fluid dispensing precision depends on the viscosity of the fluid,
its elasticity, and the total stiffness of the deposition system (notably the tubing and syringe).
The dispensing precision 
is hence estimated from the standard deviation in droplets' volume
(measured for dispensing without retraction);
we find a relative standard error of 4\% of the deposited volume.
The nozzle diameter determines the 
minimum droplet size of the droplet: 
droplets of characteristic length $\mathcal{L} \lesssim d_o $ 
would be dominated by surface interaction with the nozzle.
For a nozzle outer diameter $d_o = \SI{1.2}{\mm}$,
this amounts to a minimum droplet volume of 
$V_{min} \approx V(\mathcal{L} = 2 d_0) = 0.058 \pm 0.002$\,\si{\milli\liter}.

Each deposition system is equipped with
a nozzle that can be precisely adjusted
with a linear stage (Thorlabs DT12/M) to be at fixed distance from
the substrate.
In the current setup,
we use commercially available nozzles 
with inner and outer diameters of respectively 
\SI{0.8}{\mm} and \SI{1.2}{\mm}
(model \enquote{dispensing tip full metal single capillary}
from Vieweg).
The pre-wetting system is connected to a custom-made combination of
a multi-nozzle element and a wiper, which allows
the creation of a thin fluid layer covering the substrate (see Sec.~\ref{subsec:pre-wetting_mechanism}).

The general experimental procedure consists of:
(1)~the deposition of the precursor layer (which should happen in microgravity to ensure homogeneity and reproducible height of the layer),
followed by (2)~the droplet deposition.
The temperature inside the capsule is monitored throughout the experiment
by a type~K thermocouple,
placed on the experiment platform close to one of the deposition nozzles
(see Fig.~\ref{fig:camera_path}).
The current setup is intended for fluids whose rheology is
only weakly dependent on temperature around ambient conditions,
so that no active temperature control is needed.

\subsection{Image capture}

\begin{figure}
	\centering
    \includegraphics{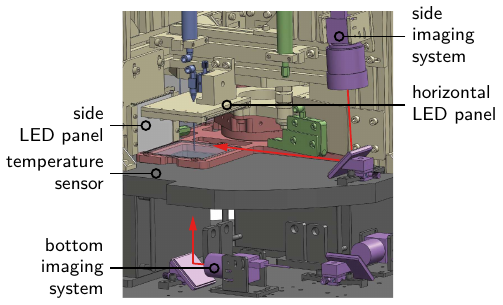}
\caption{\label{fig:camera_path}%
  Schematic of the image recording in one experiment section:
  camera heads for side- and bottom-view images are placed vertically
  above, and below the glass substrate. Camera view is indicated by
  red arrows. LED panels to the side and
  above the deposition areas provide backlight.}
\end{figure}

Different and independent measurements can be conducted simultaneously on each of the four positions.
Four cut-outs are made in the main experiment platform, allowing to
observe the droplets from below.
Therefore,
two imaging perspectives are available per position,
providing images
from the side and from below the droplet,
as visible in Figure~\ref{fig:camera_path}.
Corresponding custom-made LED panels
(\SI{4000}{\kelvin} neutral white LEDs) are placed behind and above the droplets
for backlighting, necessary for shadowgraphy and rheo-optical measurements.
The LED panel placed above the droplet 
is pierced to let the nozzle through.
For \gls{piv}, backlight is not necessary. 
For recording stress-induced birefringence patterns, the light emitted by the LED panels is polarized by a circular polarizer (Edmund Optics CP42HE) placed in front of the LED panel.
Circular polarizers are also placed on the cameras used for rheo-optical measurement.

High speed cameras are used on all sections. In the current implementation,
two Photron FastCam MC-2 recording systems available at \gls{zarm} are used,
providing four camera heads in total, each with a
\SI{512 x 512}{\pixel} CMOS chip recording at \SI{500}{fps}.
Shadowgraphy uses grayscale images, while rheo-optical measurements
use 32-bit color mode.
Two additional high speed cameras (Ximea MQ013MG-ON)
are used, which have a resolution
of \SI{1264 x 1016}{\pixel} at up to \SI{150}{fps} in color mode,
and \SI{640 x 512}{\pixel} at up to \SI{500}{fps} in grayscale. 
Standard lenses are used to obtain a resolution of around $\SI{12}{\pixel\per\milli\meter}$ respectively (for the Ximea cameras) $\SI{20}{\pixel\per\milli\meter}$.

\subsection{Particle image velocimetry}

\begin{figure*}
	\centering
    \includegraphics{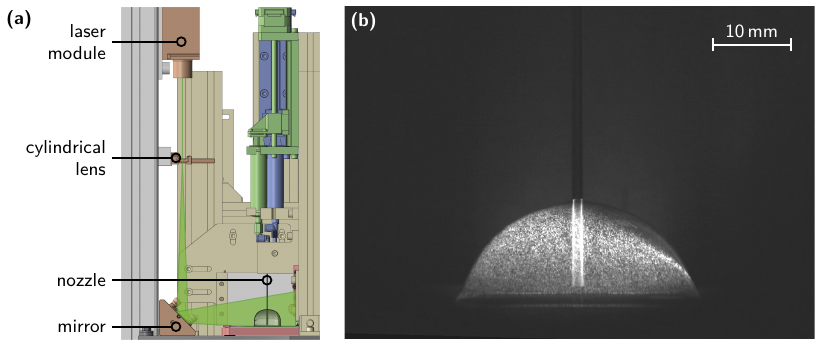}
	\caption{\label{fig:laser_path}
  Laser setup and beam guidance. (a) Illustration of the setup,
  showing the vertically mounted laser unit, the light-sheet optics
  to the side of the deposition system, together with a deposited droplet on the substrate. 
  (b) Exemplary camera recording from the side of the droplet. In this image, the laser is coming from the left side of the droplet, as pictured in (a).
}
\end{figure*}

The laser used for \gls{piv} is
mounted vertically on the main stringer structure, oriented downwards,
as shown in Figure~\ref{fig:laser_path}a.
The laser used is a \SI{532}{\nm} \gls{cw} \gls{dpss} laser
(LaserGlow LRS-0532) with optical output power of \SI{500}{\milli\watt}.
A laser sheet is formed from the laser beam using a cylindrical lens
(Edmund \SI{12.7 x 12.7}{\mm}, \SI{-25}{\mm} FL, Uncoated Laser Grade PCV),
and oriented across the droplet using a \ang{45} mirror 
(Thorlabs Broadband Dielectric Mirror, \SIrange{400}{750}{\nm}).
An optical filter (Thorlabs \SI{625}{\nm} CWL, Hard Coated OD 4.0, 
\SI{50}{\nm} Bandpass Filter) is placed on the side-view camera objective
to filter out the wavelengths captured besides those emitted by the 
fluorescent tracer particles,
resulting in images similar to the one shown in Figure~\ref{fig:laser_path}b.

For additional information on the experimental procedure,
three diagnostic cameras (GoPro Hero10) are installed on the main experiment
platform. 
Two are placed to visualize two opposite positions on the setup, while one records the entire experimental platform from above.

\subsection{Pre-wetting mechanism}\label{subsec:pre-wetting_mechanism}

The droplet deposition can be done directly on the substrate or
preceded by the deposition of a precursor layer,
constituted of either the same or a different fluid.
If the film is composed of the same fluid, 
the analysis can abstract from complicated surface-specific interactions.
Under these conditions, a droplet of a Newtonian fluid would evolve from the initial shape toward a completely flat state
\cite{Jalaal2019b,Bergemann2018}.
Droplets of yield-stress fluids, on the other hand, evolve with time towards a
final shape of finite extent,
where a balance between surface tension, the material's
yield stress and hydrostatic pressure (in the presence of gravity) is
attained~\cite{Jalaal2020a}.

While the precursor layer simplifies the physics of the problem and the associated theory
by abstracting from surface-chemistry effects,
it poses an experimental challenge: a thin film of the yield-stress fluid
needs to be created reproducibly directly before each experiment. In the
case of experiments in weightlessness, this typically needs to be attained
by an automated procedure just at the beginning of the microgravity phase,
to avoid any evaporation effects that would take place
over the prolonged preparation time of a microgravity experiment
(up to hours), and also since the films cannot be expected to be stable
with respect to the strong acceleration at launch.

An automated system allows to deposit a precursor layer on the substrates, 
prior to the droplet deposition,
by depositing a small quantity of material (in form of an array of droplets)
and spreading it
into a homogeneous film.
As mentioned previously,
the syringe used for pre-wetting procedure is distinct
from that used for the droplet deposition,
allowing the precursor layer 
to be of a different fluid
from that of the main droplet.

\begin{figure*}
	\centering
    \includegraphics{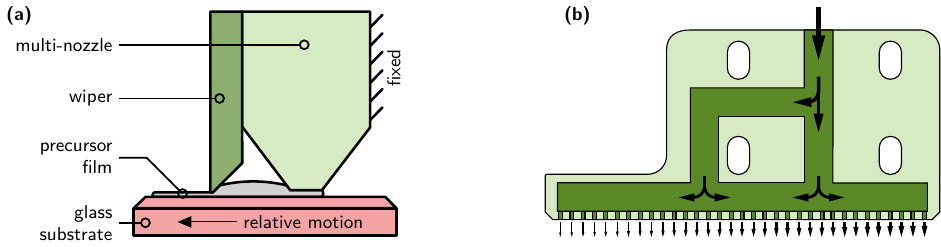}
	\caption{(a)~Schematic side-view of the pre-wetting mechanism, consisting of the deposition of a precursor layer. The multi-nozzle is used to deposit fluid on the glass substrate, which is spread into a thin film by the wiper once the glass substrate is moved below it. (b)~Multi-nozzle section cut. Arrows indicate the flow direction through the internal channels. The diameters of the outlets decrease towards the center of rotation to equalize the amount of material deposited.}
	\label{fig:pre-wetting-mechanism}
\end{figure*}

The pre-wetting mechanism, shown in Figure~\ref{fig:pre-wetting-mechanism},
consists of the assembly of a custom-made array of nozzles (labelled \emph{multi-nozzle} and manufactured by 3D printing),
a fixed steel blade 
chamfered at \ang{45}
to create a defined edge (labelled \emph{wiper}),
and a glass substrate moving under the wiper as the rotary stage rotates.
The multi-nozzle first deposits an array of droplets over its full length,
which are then mechanically spread as the glass substrate moves under the wiper.

\begin{figure*}
\centering
\includegraphics{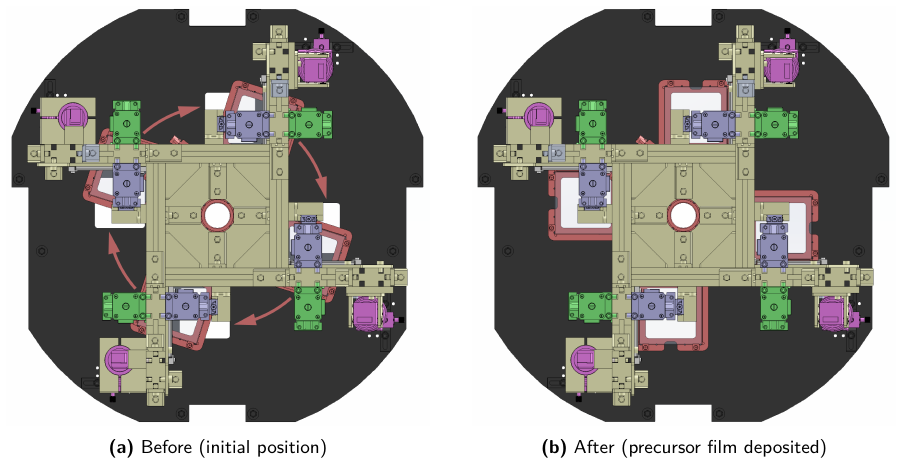}
\caption{Rotary stage motion during the experiment: (a)~Initial position, where the glass substrates are placed right before the wipers; a clockwise rotation of the stage (indicated by arrows) moves the glass substrates underneath the stationary wipers.
(b)~Final experiment position, where the substrates are centered under the deposition nozzles.}
\label{fig:rotary_plate_movement}
\end{figure*}

All four glass substrates are fixed to a rotary stage (Zaber X-RSB120AD), 
which rotates around the center of the main experiment platform. 
With this, it is possible to execute the pre-wetting procedure on all four positions simultaneously,
as the stage rotates by $\approx \SI{80}{\degree}$ clockwise,
as depicted in Figure~\ref{fig:rotary_plate_movement}.

\begin{figure*}
	\centering
    \includegraphics{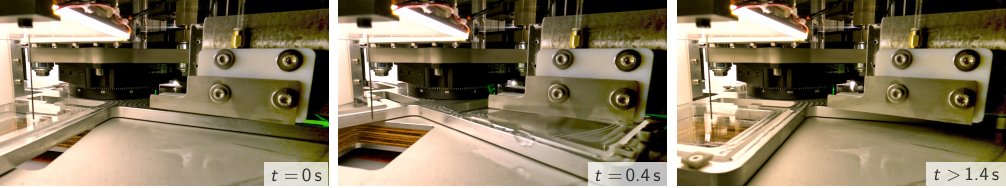}
	\caption{\label{fig:prewet_photos_seq}
Motion of the rotary stage during the pre-wetting procedure. From left to right:
initial pre-wetting position; during rotation; final position under the deposition nozzle.
The images are taken from one of the diagnostic cameras installed on the setup.
	}
\end{figure*}

The full pre-wetting procedure takes approximately \SI{1.4}{\s} and is executed 
after the hypergravity phase of the capsule launch.
A sequence of snapshots is shown in Figure~\ref{fig:prewet_photos_seq}:
initially, the rotary stage is positioned such that the inner
tip of the blade is in front of the edge of the
substrate, ensuring that fluid is spread over the entire glass
surface.
Given that the arc length of the path described by the wiper blade on the
glass substrate increases with increasing distance to the center of
rotation, the area on the substrate that needs to be covered by each segment
of the wiper also increases with radial position.
To homogenize the amount of material spread on the substrate,
the multi-nozzle is designed to deposit a radially varying amount of fluid
(less on the inner radius, more on the outer radius of the rotary stage)
by having increasingly large outlets towards the outer radius (see
Fig.~\ref{fig:pre-wetting-mechanism}b).

The homogeneity of the fluid layer depends on multiple factors,
including the rheology of the material deposited,
the timing and speed of successive events 
(e.g., onset of deposition of pre-wetting fluid and rotation of the rotary stage),
the angle between the wiper and the glass substrate,
and the quantity of material deposited by the multi-nozzle.
The parameters used for pre-wetting are the following: 
extrusion flow rate of \SI{4.2}{\milli\liter\per\second},
rotary stage rotation speed of \SI{101}{\degree\per\second},
wiper angle of \SI{45}{\degree} and multi-nozzle internal channel geometry as shown in Fig.~\ref{fig:pre-wetting-mechanism}b.

\begin{figure}
\centering
    \includegraphics{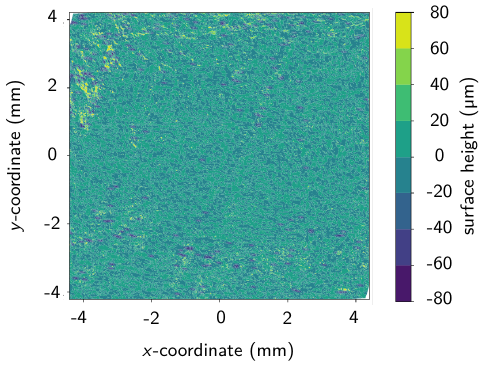}
\caption{\label{fig:pre-wetting-testing} 
\Acrfull{oct} analysis of a
precursor layer deposited through the automated pre-wetting mechanism. Height variations are measured over
an area of \SI{8 x 8}{\milli\meter} localized in the center of the glass substrate.}
\end{figure}

To test the pre-wetting system,
the height variations of an exemplary layer have been probed 
by \gls{oct} (Thorlabs Telesto Series SD-OCT Systems)
for an aqueous solution of Carbopol of 0.55 wt.\%
(fluid c.6 in Tab.~\ref{tab:parameters_carbopol}).
The result is shown in Figure~\ref{fig:pre-wetting-testing} over an area of \SI{8 x 8}{\milli\meter} localized in the center of the glass substrate:
a resolution of approx.~$\pm$~\SI{50}{\micro\meter}
is achieved for the precursor layer.

The patterns visible in Fig.~\ref{fig:pre-wetting-testing} most likely
originate from partial drying occurring on the fluid layer
between deposition and testing (approx.~\SI{10}{\minute}); 
during experiment however, the droplet deposition starts instantaneously after pre-wetting.

\subsection{Microgravity considerations}

\begin{figure*}
	\centering
    \includegraphics{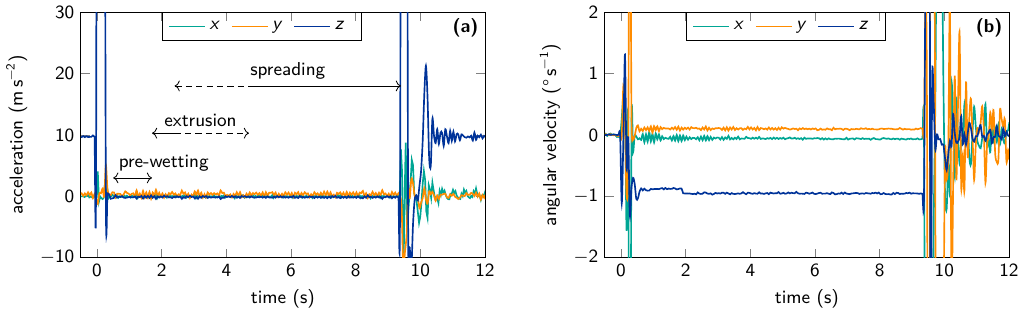}
\caption{\label{fig:accel}
Accelerometer and gyroscope data, recorded by the internal measurement unit placed inside the capsule during catapult launch, where the launch happens at $\mathsf{t_0} = \SI{0}{\second}$. (a)~Acceleration of the capsule
and (b)~angular velocity
in Cartesian coordinate directions.
The steps of the experiment and their duration are given in (a); the duration of extrusion varies with droplet size and extrusion speed: extreme values are represented by the start and end of dashed line).
}
\end{figure*}

Objects of a certain mass in motion during microgravity 
prompt the need for an analysis of the
residual acceleration and angular velocities of the capsule.
Figure~\ref{fig:accel} presents the acceleration profile (Fig.~\ref{fig:accel}a) and angular velocity (Fig.~\ref{fig:accel}b) for one catapult experiment, 
where launch happens at $t_0 =\SI{0}{\s}$,
and the capsule reaches the deceleration pit at $t \approx \SI{9.3}{\s}$.
A standard accelerometer was used to obtain the acceleration
(resolution around $10^{-2} g$),
hence the precise residual acceleration in microgravity cannot be obtained from this measurement.

Note that the overall volume of fluid that is
deposited (on the order of \SI{20}{\milli\liter}) is small enough that
no influence on the $\mu g$ quality is expected.
However, 
the rotary stage motion causes a counter-rotation of the
entire capsule due to conservation of angular momentum. 
This is visible in the angular velocity around the $z$-axis
(Fig.~\ref{fig:accel}b):
the $z$-axis angular velocity shows a small plateau lasting
until $t\approx\SI{2}{\second}$,
which corresponds to the end of the rotary stage motion. After that,
the angular velocity remains at a stable small value
(around $\SI{0.02}{\radian\per\second}$) typical for catapult launches
\cite{Selig2010}).

\section{Results}\label{sec:results}

The \textsc{vip-drop\textsuperscript{2}} module 
was used on three 
campaigns at \gls{zarm} between November 2021 and June 2022,
hence producing data both on-ground and in microgravity,
using the drop tower's catapult system.
We report here preliminary data analysis to exemplify the capabilities
of the module. A more in-depth discussion of physical effects will be
provided elsewhere.
Data obtained under microgravity conditions is labeled
$\mu g$, and corresponding ground experiment is labeled $1g$.

Two material classes are used during experiments. Shadowgraphy and
\gls{piv} are performed on aqueous solution of \gls{paa}
(referred to in the following under its commercial name Carbopol). 
The stress-optical measurements are performed on micellar solutions
of \gls{ctab} and \gls{nasal} at varying concentrations, keeping the
\gls{ctab}/\gls{nasal} ratio fixed at $5/3$.

\subsection{Experimental fluids: Carbopol solutions} 

Carbopol aqueous solutions are 
widely studied as model viscoplastic fluids with well characterized
bulk rheology (see among others Refs.~\cite{Jalaal2020a, bonn2017yield, frigaard2019simple, jalaal2019viscoplastic, Martouzet2021}).
The solutions are prepared from a 
master batch (\gls{paa} from Sigma Aldrich mixed in MilliQ water),
diluted
into six different concentrations, providing fluids with
yield stresses $\tau_0 \in [ \SI{6.5}{\pascal} , \, \SI{55}{\pascal} ]$. 
The mixing procedure is detailed in supplementary material (Sec.~\ref{sec:carbopol_protocol}).

The specific parameters of the six solutions are given in Table~\ref{tab:parameters_carbopol}; 
the rheological properties of the fluids were obtained
using an Anton Paar rheometer (MCR 502). Shear rheology (varying the shear rate
within $[\SI{0.01}{\per\second}, \, \SI{1000}{\per\second} ]$ were performed and the values of yield
stress were obtained by fitting a Herschel-Bulkley fit through the flow curve
data (Tab.~\ref{tab:parameters_carbopol}).

The surface tension of yield stress fluids is challenging to determine experimentally~\cite{Jorgensen2015, Boujlel2013}. 
For Carbopol, we use the surface tension of water ($\sigma \approx\SI{0.07}{\newton\per\meter}$)~\cite{Jorgensen2015, Jalaal2020a}

\begin{table}[h!]
\caption{ \label{tab:parameters_carbopol}	 Experimental parameters for the Carbopol aqueous solutions used in experiments.
}
    \centering
    \small 
\renewcommand{\arraystretch}{1.15} 
    \begin{ruledtabular}
     \begin{tabular}{c c c} 
	{ Fluid  } & { Composition }  & {Yield stress } \\
\hline
	Fluid c.1	&	0.3~wt\%	&	\SI{6.5}{\pascal}	\\	
	Fluid c.2	&	0.35~wt\%	&	\SI{9}{\pascal}			\\	
	Fluid c.3	&	0.4~wt\%	&	\SI{14}{\pascal}		\\	
	Fluid c.4	&	0.45~wt\%	&	\SI{21}{\pascal}				\\	
	Fluid c.5	&	0.5~wt\%	&	\SI{35}{\pascal}	\\
	Fluid c.6	&	0.55~wt\%	&	\SI{55}{\pascal}		\\
    \end{tabular}
    \end{ruledtabular}
\end{table}

\subsection{Experimental fluids: micellar solutions} 

The stress-optical measurements are performed on micellar solutions
of \gls{ctab} and \gls{nasal} at varying concentrations, keeping the
\gls{ctab}/\gls{nasal} ratio fixed at $5/3$.
These micellar solutions are exemplary shear-thinning
viscoelastic fluids that exhibit a large photoelastic constant $C$.
The stress-optical coefficient $C$ depends only on the local
structure of the polymer~\cite{DoiEdwards} and has indeed been
found to be essentially independent on the \gls{ctab}/\gls{nasal}
concentration,
$C \approx -3.1 \cdot 10^{-7} $~\si{\per\pascal}~\cite{Shikata1994}.
We neglect small nonlinearities that are in principle
present at high flow rates~\cite{Pathak2006}.

Using \gls{ctab}/\gls{nasal} solutions of the same mixing ratio,
\citeauthor{Gladden2007}
demonstrated rheo-optical measurements of the stresses around a moving object
in a study of cutting and tearing of gel \cite{Gladden2007} and of
shear waves in the fluid \cite{Gladden2010}; an equimolar mixture
of \gls{cpcl} and \gls{nasal} was used by \citeauthor{Handzy2004}~\cite{Handzy2004}
to visualize the stresses around rising bubbles in a viscoelastic fluid.

Changing the concentration of both \gls{ctab} and \gls{nasal}, we adjust
the viscoelasticity of the solutions~\cite{Hartmann.1997}.
The fixed mixing ratio ensures
that the fluid remains in the homogeneously flowing regime of wormlike
micelles~\cite{Recktenwald.2019}.
The parameters of the stress-birefringent fluids used in
experiments are summarized in Table~\ref{tab:parameters_sb}.

\begin{table*}
 \caption{Parameters of the micellar solutions used as stress birefringent materials in experiments. Values for viscosity, structural relaxation time and apparent yield stress are estimated from literature:
$^a$ \Citeauthor{Gladden2007} \cite{Gladden2007};
$^b$ \Citeauthor{Gladden2010} \cite{Gladden2010};
$^c$ \Citeauthor{Hartmann.1997} \cite{Hartmann.1997}.
    \label{tab:parameters_sb}
}
    \centering
    \small 
    \renewcommand{\arraystretch}{1.15} 
    \begin{ruledtabular}
     \begin{tabular*}{\textwidth}{c @{\extracolsep{\fill}} c c c c} 
	{Fluid} & {Composition (CTAB/NaSal)}  & {Viscosity} & {Relaxation time } & {Apparent yield stress}  \\
	\hline
	Fluid sb.1	&	200/120~\si{\milli\Molar}	&	\SI{167}{\pascal\second}~$^a$ & 
	\SI{1}{\second}~$^b$ & \SI{150}{\pascal}~$^b$	\\
	Fluid sb.2	&	160/96~\si{\milli\Molar}	&	--\SI{}{}	& -- & -- \\
	Fluid sb.3	&	120/72~\si{\milli\Molar}&	--\SI{}{}	& -- & --\\
	Fluid sb.4	&	100/60~\si{\milli\Molar}&	\SI{1200}{\pascal\second}~$^c$ 
	& \SI{30}{\second}~$^c$	& \SI{40}{\pascal}~$^c$ \\
	Fluid sb.5	&	80/48~\si{\milli\Molar}&--	 & -- 	& -- \\
    \end{tabular*}
    \end{ruledtabular}
\end{table*}

\subsection{Experimental parameters}

Droplets of various sizes are deposited, characterized by the length scale $\mathcal{L}$, which corresponds to the radius of a perfect sphere of the same volume as the droplet deposited -- in other words, for a droplet of volume $V$, $\mathcal{L} = { (3 V / 4 \pi )}^{1/3}$.
The rheology of the experimental fluids is characterized by the plastocapillary number 
\begin{equation}
	\mathcal{J} = \frac{\tau_0 \, \mathcal{L} }{\sigma},
\end{equation}
which compares the yield stress $\tau_0$ of the fluid to its capillary pressure.
The extrusion speed at which droplets are deposited can also be varied;
the parameters that we have tested
are summarized in Table~\ref{tab:exp_parameters}.
It is noteworthy that a short phase of retraction after the deposition 
can be implemented to ensure full arrest of the extrusion.

\begin{table}
    \centering
    \caption{\label{tab:exp_parameters} Extrusion flow rates used in the \textsc{vip-drop}$^2$ module during experimental campaigns. }
\renewcommand{\arraystretch}{1.2}
\small 
\begin{ruledtabular}
\begin{tabular}{c} 
{Extrusion flow rates} \\
\hline
$Q_0 = 5.97 \cdot 10^{-1}$~\si{\milli\liter\per\second} \\
$Q_1 = 2.39 $~\si{\milli\liter\per\second} \\
$Q_2 = 4.77 $~\si{\milli\liter\per\second} \\
\end{tabular}
\end{ruledtabular}
\end{table}

Finally, the Bond number defined in Eq.~\ref{eq:bo} is used to characterize the regime in which each experiment is conducted. 
For experiments conducted under microgravity, $B$ is calculated using $g_{\mu g} = \SI{9.81e-4}{\meter\per\second\square}$,
which corresponds to the average gravitational acceleration during catapult experiments. 
The volume $V$ of each droplet (hence its true characteristic length) is calculated from the final shape of the droplet by image analysis, assuming axisymmetry.

\subsection{Shadowgraphy}

\begin{figure*}
\centering
\includegraphics{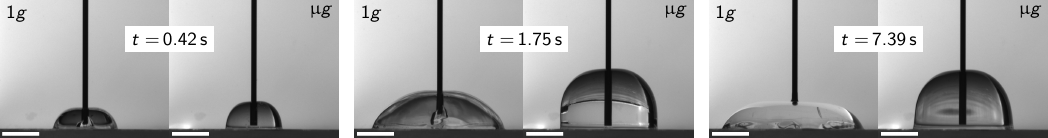}
\caption{\label{fig:droplets}
Exemplary shadowgraphy images for the carbopol dispersion c.5 (see
Table~\ref{tab:parameters_carbopol}),
on ground ($1 g$) and under microgravity conditions ($\mu g$).
The droplet has a final volume $V=$~\SI{2.7}{\milli\liter}.
The scale bar on each picture represents \SI{8}{\milli\meter}.
}
\end{figure*}

Exemplary shadowgraphy images of Carpobol solution droplets are
shown in Figure~\ref{fig:droplets}. The images are taken at various
times after the start of the fluid deposition; 
note that extrusion is not complete at the time when the first image is taken.
The images labeled $1g$ and $\mu g$ are respectively from on-ground and microgravity
repetitions of the experiment under the same parameters.
Visual inspection of the images already demonstrates the effect of
hydrostatic pressure in the $1g$ experiment, flattening the droplet
while under microgravity conditions it retains its initial
dome-like shape.

\begin{figure*}
\centering
 \includegraphics{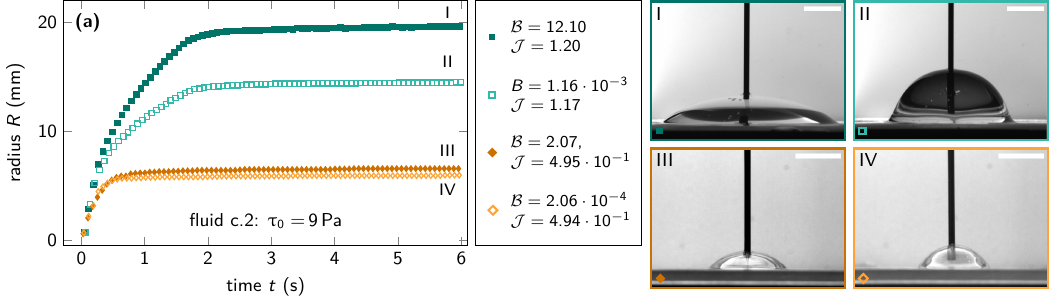}
\includegraphics{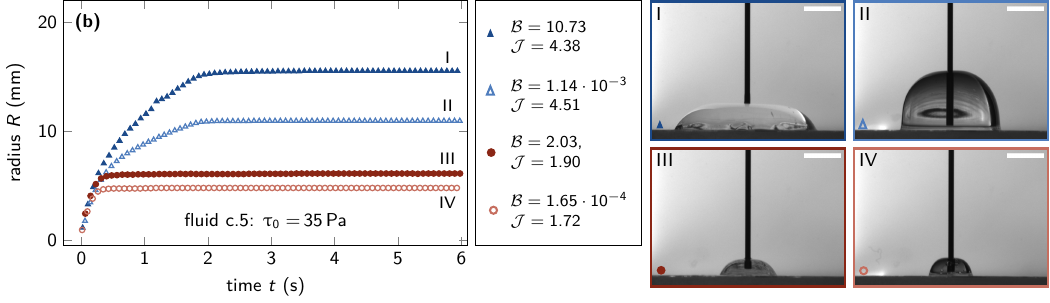}
\caption{\label{fig:radii}
Radius $R(t)$ of the droplets as a function of time $t$, comparing the
same fluids on-ground (labelled $1g$, filled marks) and under microgravity conditions
($\mu g$, empty marks).
(a)~For fluid c.2, aqueous solution of 0.5~wt\% Carbopol,
and (b)~for fluid c.5, aqueous solution of 0.35~wt\% Carbopol,
for two droplets of characteristic lengths 
$\mathcal{L}\approx 9.43 \pm 0.28$\,\si{\mm} and 
$\mathcal{L}\approx 3.92 \pm 0.25 $\,\si{\mm},
respectively for (a) green squares and orange diamonds,
and for (b) blue triangles and red circles.
Pictures are taken at the last instant before the end of the experiment.
In all pictures, the scale bar represents \SI{8}{\mm}.
}
\end{figure*}

The simplest quantity to extract from the shadowgraphy experiment is the radius of a
droplet as a function of time, $R(t)$, easily accessible by image analysis.
Figure~\ref{fig:radii} shows exemplary evolutions of $R(t)$,
where $t = \SI{0}{\s}$ is the time at which extrusion effectively starts.
The droplets' volume is estimated using the
\texttt{Sessile Drop Analysis} Python algorithm 
by~\citeauthor{vanGorcum2022}~\cite{vanGorcum2020, vanGorcum2022}.
Eight experiments are represented:
in Fig.~\ref{fig:radii}a, droplets of fluid c.2 ($\tau_0=\SI{9}{\pascal}$) are analyzed;
in Fig.~\ref{fig:radii}b, data corresponding to fluid c.5 ($\tau_0=\SI{35}{\pascal}$) is shown.
The four curves in each graph correspond to two droplet sizes (characteristic lengths $\mathcal{L}$ of approx.~\SI{4.5}{\mm} and \SI{10}{\mm}; orange respectively green symbols in Fig.~\ref{fig:radii}a, and
red respectively blue symbols in Fig.~\ref{fig:radii}b), under Earth gravity and microgravity (full and empty marks, respectively).
Besides the graphs, Fig.~\ref{fig:radii} shows images of the analyzed droplets, at a time well inside the stationary regime of $R(t)$.

The data demonstrate the influence of gravity on the droplet spreading
clearly, where in both cases, the fluid spreads faster in the presence of
gravitationally-induced hydrostatic pressure. 

\begin{figure*}
\centering
\includegraphics{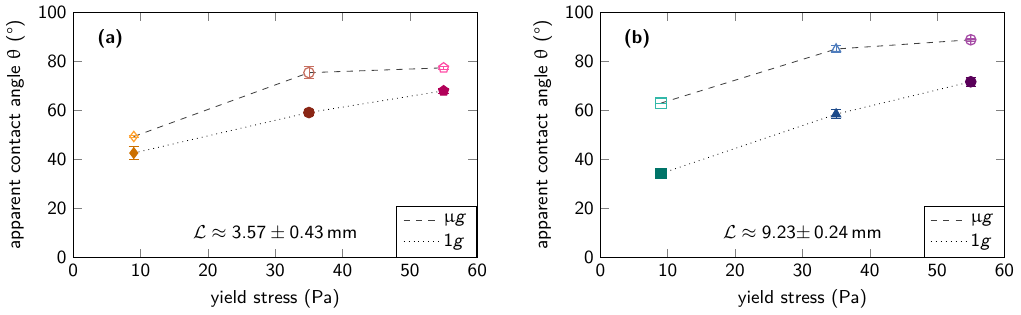}
\caption{\label{fig:contact angle}
Apparent contact angle $\theta$ between the droplets and the precursor layer, for droplets of characteristic length 
(a)~$\mathcal{L} \approx 3.51 \pm 0.43 $\,\si{\mm} and
(b)~$\mathcal{L} \approx 11.16 \pm 2.13$\,\si{\mm}.
Lines are a help to the eye, 
distinguishing $\mu g$ (dashed line, open symbols) and $1g$ (dotted line, filled symbols) data sets. Error bars are the standard deviation on measured $\theta$.
}
\end{figure*}

Because for a yield stress fluid, the spreading stops at a finite droplet radius, there results a finite apparent macroscopic contact angle between the surface-wetting layer and the droplet, even if both are of the same fluid (and furnish zero microscopic contact angle).
The apparent contact angle $\theta$ between the droplet and the precursor layer depends on the material properties, and it is also modified by the gravitational environment. 
We have extracted $\theta$ by image analysis~\cite{vanGorcum2020, vanGorcum2022} from the shadowgraphy images presented in Fig.~\ref{fig:radii}a and~\ref{fig:radii}b as I to IV: from local linear fits to the droplet profile, the value corresponding to the maximum slope is taken as the apparent contact angle. These values are reported in
Figure~\ref{fig:contact angle}.

The increase of apparent contact angle $\theta$ with droplets' radius and yield stress
is in accordance with previous work~\cite{Martouzet2021};
our results allow to consider additionally the gravitational acceleration as a variable parameter. 
Again, the effect of gravity is evident:
hydrostatic pressure causes a decrease in the contact angle,
which is more pronounced for droplets of large characteristic length (Fig.~\ref{fig:contact angle}b).

\subsection{\Acrfull{piv}}

For \gls{piv}, some of the carbopol solutions were seeded with
fluorescent-labeled tracer particles. The tracers are
\SI{10}{\micro\meter} diameter
\gls{ps} particles
from MicroParticles GmbH (product PS-FluoRed-10.0).
Polymerized with organic 
fluorescent dyes,
they absorb light at a wavelength around the laser excitation
of \SI{532}{\nano\meter}, and emit fluorescence light at
a wave length of \SI{607}{\nano\meter}.

\begin{figure*}
\centering
    \includegraphics{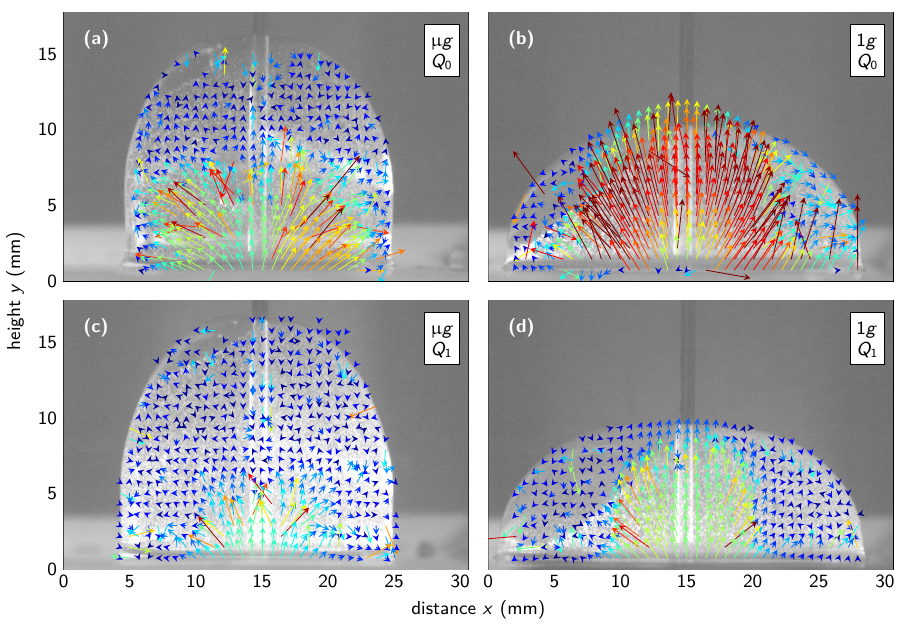}
\caption{\label{fig:piv}
  Exemplary \gls{piv} reconstructions of the velocity fields inside
  carbopol droplets (material c.6, characteristic length
  $\mathcal L\approx\SI{10}{\milli\meter}$), $t=\SI{2}{\second}$ after
  start of the droplet deposition.
  The top panels (a) and (b) correspond to an extrusion speed
  $Q_1=\SI{2.39}{\milli\liter\per\second}$ in $\mu g$ and $1g$ conditions,
  respectively; bottom panels (c) and (d) to
  $Q_2=\SI{4.77}{\milli\liter\per\second}$.
  The arrows indicating the in-plane velocity field are colored according
  to increasing magnitude (blue to red in the range $v=\SI{0}{\milli\meter\per\second}$ to
  $v=\SI{3}{\milli\meter\per\second}$).
Droplet images are to scale within the plots' axes. 
}
\end{figure*}

Exemplary velocity-field reconstructions for a point $t=\SI{2}{\second}$
at the end of the extrusion phase
are shown in Figure~\ref{fig:piv}.
The droplet side view was recorded at \SI{50}{\fps}, and \gls{piv}
was performed using two consecutive frames using the software
\texttt{OpenPIV} (version~0.23.8)~\cite{OpenPIV}.
Images were masked with a threshold map to darken the area outside the
droplet.
The snapshots in Figure~\ref{fig:piv} feature the moment at the end of the extrusion phase when the droplets spreading is strongly slowed down, and the pump is about to stop; hence, a localized flow near the nozzle appears.

The velocity fields under $1g$ show the appearance of a 
region with a significantly smaller velocity 
on the outer surface of the droplets, 
separated from the inner moving part of the fluid by a
sharply defined yield surface.
For the flatter droplet extruded
at a higher speed (Fig.~\ref{fig:piv} bottom), the location of this yield surface
is closer to the center of the droplet.

A striking difference is seen in comparison to the droplets deposited
in microgravity (Fig.~\ref{fig:piv} left): here, the zero-velocity regime
covers the entire top of the droplet. The absence of hydrostatic pressure
allows the unyielded fluid to remain on top of the droplet, without being
pushed to the sides as it is the case in $1g$.

Note that for flat droplets with a large bottom area, performing \gls{piv} from the side
has the drawback that due to a \enquote{lens effect} caused by
total inner reflection, dark regions appear in some of the
\gls{piv} images. This is partially mitigated by placing the laser sheet
slightly less centered for large droplets (to avoid crossing the nozzle).
It limits the application of \gls{piv} to fluids of sufficiently high
yield stress, where the resulting droplet shapes are more less flat.

Also visible in Fig.~\ref{fig:piv} is a \enquote{shadow} effect to the
left of the deposition needle in the top part of the droplets. This
shadow is caused by the needle, and limits the information that can be obtained
from the left half of the droplet. However, as seen in Fig.~\ref{fig:piv}(b),
even in the shadow region a reconstruction of the velocity field is still
possible within reasonable limitations.

\subsection{Rheo-optical measurement}

A detailed analysis of the rheo-optical data obtained from our setup
is beyond the scope of the present contribution. We only show here
a set of exemplary images to allow for a qualitative assessment.

\begin{figure*}
	\centering
	    \includegraphics{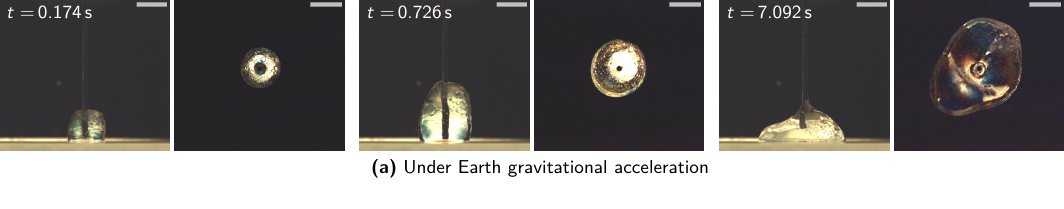}
	    \includegraphics{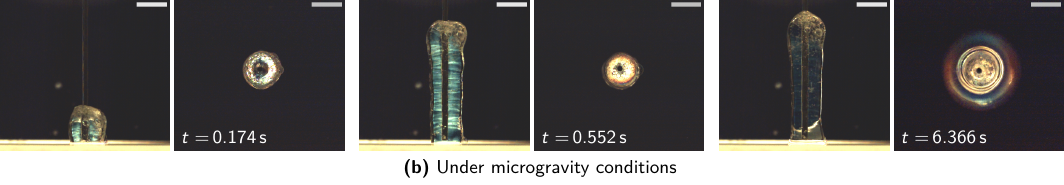}
		\vspace{-7mm}
	\caption{\label{fig:SB_fastextrusion} Droplets of micellar fluid exhibiting stress birefringence patterns, (a)~on-ground and (b)~under microgravity conditions, for the fluid sb.~1 (CTAB/NaSal 200/120~\si{\milli\Molar}).
		On both levels of gravitational acceleration, material extrusion was done at \SI{7.5e-3}{\meter\per\second}.
		Time $t$ is given as time since $t_0 =\SI{0}{\s}$ the effective start of extrusion.
		The scale bar on each image represents \SI{8}{\milli\meter}.}
\end{figure*}

In Figure~\ref{fig:SB_fastextrusion},
droplets of the micellar fluid sb.\,1 (CTAB/NaSal 200/120~\si{\milli\Molar})
are shown at three representative instants in the experiment:
first,
shortly after the effective start of extrusion;
then right before the end of extrusion;
finally, in the last instant preceding 
impact for the microgravity case, and at the exact same time for the ground one.
The fluid is deposited at an extrusion flow rate of
$7.5 \cdot 10^{-3}$~\si{\meter\per\second}.
In all pictures presented, the droplets are deposited directly on the glass substrate, without precursor layer.

From the exemplary images presented, one can see that there is a dramatic
effect of the deposition rate for the micellar fluid, both on the shape
of the droplet and on the magnitude of the internal stresses.
In particular, for the faster deposition rate, we observe the formation of
a high \enquote{pillar} of the viscoelastic fluid under microgravity conditions,
that remains stable over the time scale of the experiment. The same
material, with the same extrusion speed, but on-ground, forms much flatter
droplets, demonstrating the role of microgravity conditions to stabilize
the extrusion of tall structures of a soft material.

A rough estimate of the stresses
from the blue-ish color clearly visible in fig.~\ref{fig:SB_fastextrusion}, making use of suitable interference color charts~\cite{Sorensen},
confirms that the internal stresses are comparable to the yield
stress of the material, around \SIrange{10}{50}{\pascal}.

\section{Conclusion}\label{sec:conclusion}

We have described our implementation of the \textsc{vip-drop}$^2$ drop-tower module to study
the spreading of complex fluids under microgravity.
The motivation to perform such studies is given by the fact that
droplet spreading in the presence of gravity, i.e., at finite Bond
number, is notably different from the spreading in the surface-tension
dominated regime at $\mathcal B \rightarrow 0$.
The latter regime is of interest for many technological applications,
among which \gls{am}.
Performing experiments under microgravity conditions, large droplets
can be used to carry out optical analyses of droplet shapes,
internal velocity and stress fields, while maintaining
small Bond numbers.
The microgravity experiment essentially achieves to decouple the limit
an ultimately of $\mathcal B \rightarrow 0$ from the material and deposition parameters of
the yield-stress fluids.

A more detailed analysis of the results provided by our drop-tower module
will address the comparison with scaling laws predicted by the
thin-film solutions of the Navier-Stokes equations for viscoplastic
fluids. It is expected, that the case $\mathcal B=0$ is governed by
different scaling exponents, and achieving this limit with large droplets
will allow to verify the scaling law without being hampered by experimental
artifacts arising from small droplets.

The \gls{piv} measurements can be used to determine the gravity-dependence
of the yield surface inside the droplet, separating the region of plastic flow
in the region with large stresses from the \enquote{plug} flow in the region
where the stresses remain below the yield stress.
Future extensions of the setup could include \gls{ptv}, a Lagrangian method where the tracer
particles are individually tracked. Suitable methods have been developed
for droplets, including astigmatic \gls{ptv}~\cite{Rossi2020},
which allows the tracking in 3D with a single camera, utilizing its
astigmatic imaging defects to reconstruct out-of-focal plane positions.
\Gls{oct} has also allowed to study the flow fields of evaporating droplets
at various inclination angles (and hence various effective gravity levels)~\cite{Edwards2018}, and could be added in a future implementation of the experiment.

We aim to establish the analysis of stress-induced birefringence in
droplets of yield-stress fluids (or shear-thinning viscoelastic fluids
with a large apparent yield stress) as a qualitative, and ultimately a
quantitative tool to predict internal stresses in yield-stress materials.
To this end, the birefringent images obtained in our experiment
will be compared to computer simulation combined with ray tracing,
potentially using machine learning to address the inverse problem
of reconstructing the internal stress field from the
experimental images.

For the present analysis, we have largely ignored the dynamics of the
droplet in the moment of impact at the end of the drop-tower flight.
These moments under strong hypergravity conditions could potentially
be exploited further to provide data on
the effect of \emph{external} acceleration on the fluid's spread,
such as in centrifugal casting 
or density-based separation methods.

Further developments could include an adaptation of the experimental module
for different microgravity platforms in order to provide longer time in
weightlessness, for example to study the deposition of filaments of
viscoplastic materials that are relevant for \gls{am} and bioprinting.
Possible platforms include sounding rockets, which providing several minutes
of experiment time in weightlessness.

\section*{Supplementary material}

\subsection{Mixing protocol for Carbopol solutions}\label{sec:carbopol_protocol}

For Carbopol suspensions, premixed aqueous solution of 1 wt\% of \acrfull{paa}
polymer (Sigma Aldrich) was first
prepared: \acrfull{paa} powders were mixed with Milli-Q water using a
four-blade marine impeller (1000-1500 rpm) at room temperature. The mixture was
pH-neutralized with triethanolamin (Sigma Aldrich). The premixed batch was then
diluted to the final concentrations listed in table~\ref{tab:parameters_carbopol}. Bubbles were removed from the sample by centrifuging the
fluids at \SI{2200}{\rpm} for \SI{20}{\minute}.

\subsection{Mixing protocol for micellar solutions}\label{sec:micellar_solution_protocol}

The aqueous wormlike micellar solutions
are obtained from 
\acrfull{ctab} in powder ($\geq 99\%$ pure \chem{C_{19}H_{42}BrN}, Carl Roth)
and \acrfull{nasal} ($99\%$ pure \chem{C_{7}H_{5}NaO_3}, ThermoFisher Scientific).
Each powder is individually dissolved in distilled water at \SI{70}{\celsius} and strongly stirred until full dissolution of the powder is the water (with a minimum of \SI{30}{\minute}). 
Then both fluids are mixed; the mixture is covered and strongly stirred for for \SI{1}{\hour} at \SI{70}{\celsius}, before being let to cool down for at least one day before use. 
Bubbles are suppressed by centrifuging the fluids at \SI{7000}{\rpm} for \SI{10}{\minute}.

\subsection{Videos}

Videos are provided as supplementary material to illustrate the droplet deposition under Earth gravity and microgravity conditions.
For all videos, the calibration bar corresponds to a length of \SI{8}{\mm} and the video is slowed down by a factor of two. 
The videos provided correspond to the following experiments:
\begin{itemize}[label={-}]
	\item Fluid c.1 (lowest yield stress), droplet of characteristic length 
$\mathcal{L} = 9.92  \pm 0.04$\,\si{\mm},
extrusion flow rate $Q_1 = \SI{2.39}{\milli\liter\per\second}$; on-ground (left) and in microgravity (right). 
	\item Fluid c.6 (highest yield stress), droplet of characteristic length
$\mathcal{L} = 10.30  \pm 0.11$\,\si{\mm},
extrusion flow rate $Q_1 = \SI{2.39}{\milli\liter\per\second}$; on-ground (left) and in microgravity (right). 
	\item Fluid sb.\,1 (CTAB/NaSal 200/120~\si{\milli\Molar}), droplet of characteristic length $\mathcal{L}=6.95  \pm 0.01$\,\si{\mm},
extrusion flow rate $Q_1 = \SI{2.39}{\milli\liter\per\second}$; on-ground (top, both panels) and in microgravity (bottom, both panels). For both gravitational accelerations, the two panels placed side by side are the side (left) and bottom (right) views of the same droplet.
\end{itemize}

\section*{Author declarations}
The authors have no conflicts to disclose.

\section*{Data availability}
All data is available from the corresponding author upon reasonable request.

\begin{acknowledgments}
ODA acknowledges the 
\acrfull{elgra},
who initially supported this project through the 
2021 \acrshort{elgra} Research Prize.
MJ acknowledges the funding provided by \acrfull{ixa} via the project \enquote{Sprinter} and the \acrfull{nwo} via the project 3D Printing Soft Matters in Space (XS21.1.140).
The team also acknowledges the \acrfull{esa}
for providing the drop-tower catapult opportunities
through the \acrshort{esa}-CORA program.

The personnel of the \acrfull{zarm},
and in particular 
Thorben K\"onemann,
Fred Oetken and Jan Siemer,
are warmly acknowledged for their professional yet friendly 
help during the campaigns.
The \acrfull{tc} of the University of Amsterdam has been essential in the construction of the \textsc{vip-drop}$^2$ experiment module. In particular,
Tjeerd G.L.C.~Weijers and Daan Giesen are personally acknowledged for their 
support.
ODA wishes to express her gratitude to Martin Castillo
and Cyprien Verseux
for the use of their laboratories during campaigns
and insightful discussions. 
Appreciation extends to Marion Casanova and Linnea Heitmeier.
\end{acknowledgments}

\section*{References}
\bibliography{bibliography}

\end{document}